\documentclass[12pt]{article}
\usepackage[dvips]{graphicx}
\usepackage{pdproc}
\usepackage{epsfig}
\usepackage{subfigure}
\usepackage{amssymb}
  \makeatletter 
  \def\@cite#1{[#1]} 
  \makeatother    
\def\bea{\begin{eqnarray}}
\def\eea{\end{eqnarray}}
\def\be{\begin{equation}}
\def\ee{\end{equation}}

\begin{document}

\renewcommand{\thefootnote}{\alph{footnote}}

\title{Neutrino Oscillations and Collider Test of
the R-parity Violating mSUGRA Model \footnote{To appear in the 
Proceedings of "SUSY 2004", Tsukuba, Japan, 17-23 Jung, 2004}
}

\author{ DONG-WON JUNG}

\address{ 
 Department of Physics, Korea Advanced Institute of Science and
 Technology,\\
373-1 Kusong-dong, Yusong-gu, Daejoen 305-701, Korea 
\\ {\rm E-mail: dwjung@muon.kaist.ac.kr}}

\abstract{
We study the R-parity violating minimal supergravity models accounting
for the observed neutrino masses and mixing, which can be tested in
future collider experiments. The bi-large mixing can be explained
by allowing five dominant tri-linear couplings $ \lambda'_{1,2,3}$
and $\lambda_{1,2}$.  The desired ratio of the atmospheric and
solar neutrino mass-squared differences can be obtained
in a very limited parameter space where the tree-level contribution
is tuned to be suppressed.  In this allowed region, we quantify
the correlation between the three neutrino mixing angles and
the tri-linear R-parity violating couplings.  
Such a prediction on the couplings can be tested in the next linear
colliders by observing the branching ratios of the lightest supersymmetric
particle (LSP).
For the stau or the neutralino LSP, the ratio
$|\lambda_1|^2: |\lambda_2|^2: |\lambda_1|^2 + |\lambda_2|^2$
can be measured by establishing
$Br(e\nu): Br(\mu\nu) : Br(\tau\nu)$ or
$Br(\nu e^\pm \tau^\mp ): Br(\nu\mu^\pm\tau^\mp) : Br(\nu\tau^\pm\tau^\mp)$,
respectively.
The information on the couplings $\lambda'_i$ can be
drawn by measuring $Br(l_i t \bar{b}) \propto |\lambda'_i|^2$ if
the neutralino LSP is heavier than the top quark.
}

\normalsize\baselineskip=15pt

\section{Neutino Mass Matrix}
Let us begin by writing the superpotential in
the basis where the bilinear term $L_i H_2$ is rotated away\cite{hall} :

\begin{equation}
W_0 = \mu H_1 H_2 + h^e_i L_i H_1 E^c_i + h^d_i Q_i H_1 D_i^c
 + h^u_i Q_i H_2 U^c_i ,
\end{equation}

\begin{equation}  \label{WRpV}
W_1=  \lambda_{i} L_i L_3 E^c_3 + \lambda'_{i} L_i Q_3 D^c_3,
\end{equation}
where  $W_0$ is R-parity conserving part and $W_1$ is R-parity
violating part.
Here, we have taken only 5 trilinear couplings, $\lambda_i$ and $\lambda'_i$,
assuming the usual hierarchy of Yukawa couplings.
With the non-zero sneutrino vacuum expectation values \cite{chun}
\begin{equation}\label{svev}
\xi_i \equiv \frac{\langle \tilde{\nu_i} \rangle}
{\langle H_1^0 \rangle} = - {m^2_{L_i H_1} + B_i t_\beta +
\Sigma_{L_i}^{(1)}
\over m^2_{\tilde{\nu}_i} + \Sigma_{L_i}^{(2)} },
\end{equation}
where the 1-loop contributions $\Sigma_{L_i}^{(1,2)}$ are given by
$\Sigma_{L_i}^{(1)}=\partial V_1/H_1^{\ast}\partial L_i$,
$\Sigma_{L_i}^{(2)}=\partial V_1/L_i^{\ast}\partial L_i$,
the light tree-level neutrino mass matrix
of the form ;
\begin{equation} \label{Mtree}
M^{tree}_{ij} = -{M_Z^2 \over F_N} \xi_i \xi_j \cos^2\beta ,
\end{equation}
where $F_N= M_1M_2/M_{\tilde{\gamma}}+ M_Z^2 \cos{2\beta}/\mu$ with
$M_{\tilde{\gamma}} = c_W^2 M_1 + s_W^2 M_2$.
The R-parity violating vertices between particles and sparticles
can give rise to 1-loop neutrino masses. Including all the 1-loop
corrections, the loop mass matrix can be written as \cite{CK}
\bea \label{neutrinoloop}
M^{loop}_{ij}= 
-\frac{M_Z^2}{F_N} \left( \xi_i \delta_j +\delta_i \xi_j \right)
\cos\beta +\Pi_{ij},
\eea
where $\Pi _{ij} $ denotes the 1-loop contribution of the neutrino
self energy and
\bea \label{comentary} \delta_i &=& \Pi_{\nu_i
\widetilde{B}^0} \left(
\frac{-M_2\sin^2\theta_W}{M_{\widetilde{\gamma}} M_W \tan\theta_W}
\right) + \Pi_{\nu_i \widetilde{W}_3} \left( \frac{M_1
\cos^2\theta_W}{M_{\widetilde{\gamma}}M_W} \right) \nonumber \\
&&+\Pi_{\nu_i \widetilde{H}_1^0} \left( \frac{\sin\beta}{\mu}
\right) +\Pi_{\nu_i \widetilde{H}_2^0}\left(
\frac{-\cos\beta}{\mu} \right).
\eea

\section{Numerical Analysis}
For our numerical analysis, we scan the input parameters in the following
ranges;
\bea
100 {\rm GeV} \leq &m_0& \leq 1000 {\rm GeV}, \\
100 {\rm GeV} \leq &M_{1/2}& \leq 1000 {\rm GeV},\nonumber \\
0 {\rm GeV} \leq &A_0& \leq 700 {\rm GeV},\nonumber \\
2 \leq &\tan \beta& \leq 40\nonumber .
\eea
The ranges for $R-$parity  violating parameters we scan are
\bea
4 \times 10^{-6} \leq & \left| \lambda_1 \right|& \leq 6 \times
10^{-4}, \\
4 \times 10^{-6} \leq & \left| \lambda_2 \right|&\leq 6 \times
10^{-4},\nonumber \\
3 \times 10^{-9} \leq & \left| \lambda_1^{'} \right|& \leq
10^{-4},\nonumber \\
4 \times 10^{-6} \leq & \left| \lambda_2^{'} \right|& \leq
10^{-3},\nonumber \\
4 \times 10^{-6} \leq & \left| \lambda_3^{'} \right|& \leq
10^{-3}.  \nonumber
\eea
We set the signs of $ M_{1/2}$ and $A_0$ arbitrary,
but find that most of the allowed  parameter space  corresponds to
the case that both signs are positive.
In the mSUGRA
model, the tree-level contribution to the neutrino mass matrix is
too much large compared with the one-loop contribution in the
generic parameter space.  In order to obtain the observed value of $
\Delta m^2_{sol}/\Delta m^2_{atm} $, the parameters needs to be
tuned to arrange some cancellation in the tree-level mass which is
made to be comparable to the loop mass.   As a result, a nice
relation between the R-parity (and lepton number) violating
parameters and the neutrino oscillation parameters is lost. This is
a dis-favorable feature of the mSUGRA model with R-parity violation
as the origin of the observed neutrino mass matrix. However, in  the
allowed region of the parameter space, we still find the reasonable
correlations between the tri-linear R-parity violating couplings and
the neutrino mixing angles. That is, the bi-large mixing of the
atmospheric and solar neutrino oscillation requires $|\lambda'_2|
\sim | \lambda'_3 |$  and $| \lambda_1| \sim |\lambda_2|$ ,
while the small mixing angle $\theta_{13}$  requires $|\lambda'_1|
\ll |\lambda'_{2,3}|$.
Even though $\lambda'_2/\lambda'_3$ has no 
analytic relation with $\theta_{23}$ for the solution points,
we can obtain the following favorable  ranges through
the parameter scan:
$\left|\lambda'_2/\lambda'_3 \right|$ from the neutrino data
\begin{eqnarray}
0.4 \lesssim |\lambda'_2/\lambda'_3| \lesssim 2.5 &\mbox{for}&
\tan\beta = 3-15 \nonumber \\
0.3 \lesssim |\lambda'_2/\lambda'_3| \lesssim 3.3 & \mbox{for}&
\tan\beta =  30- 40.
\end{eqnarray}
Similar to the case of the atmospheric neutrino oscillation, we can get
the constraints:
\begin{eqnarray}
0.3 \lesssim |\lambda_1/\lambda_2| \lesssim 1.6 & \mbox{for} &
\tan\beta=3-15, \nonumber \\
0.2 \lesssim |\lambda_1/\lambda_2| \lesssim 5.0 & \mbox{for} &
\tan\beta=30-40.
\end{eqnarray}

\begin{figure}
\begin{center}
\subfigure[$\sin^2 2 \theta_{12}$ vs. $|\lambda_1 / \lambda_2|$. ]
{\epsfig{file=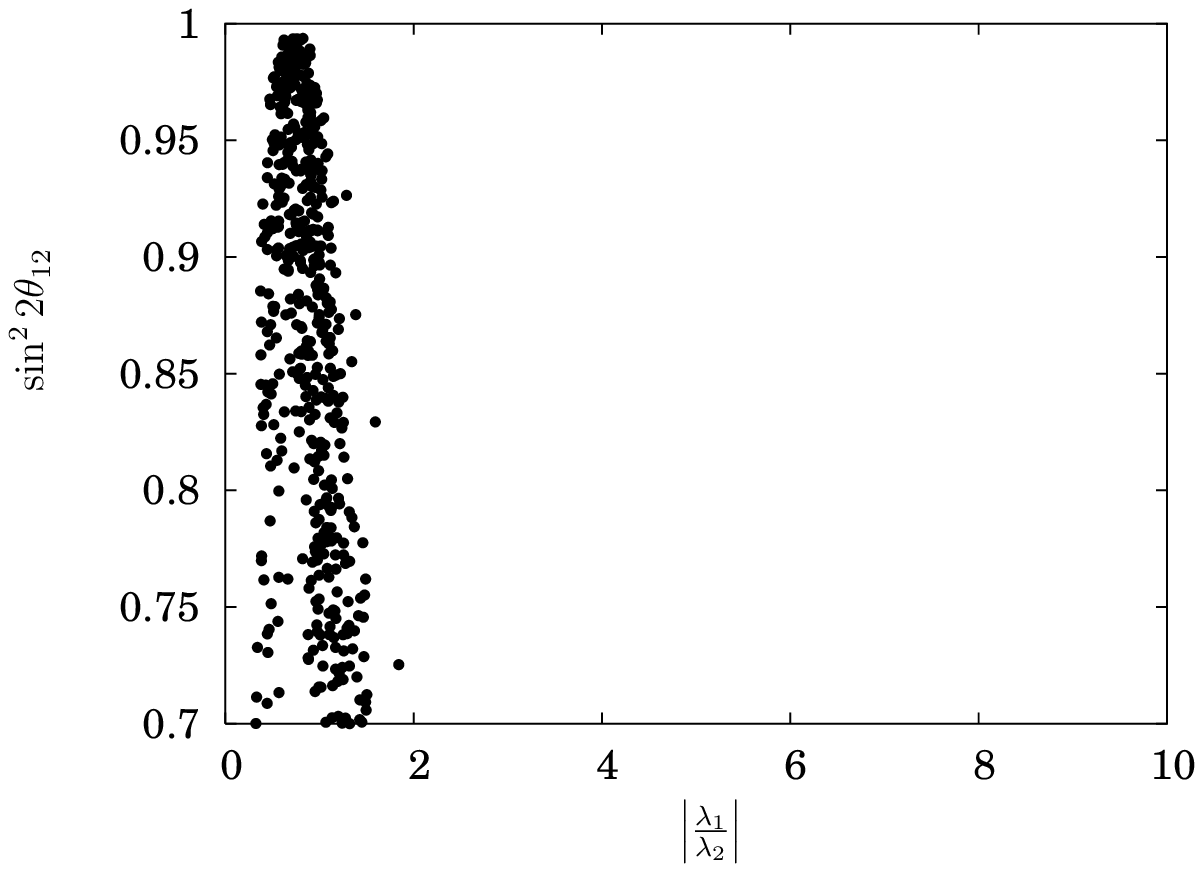,height=3.9cm,width=6.0cm}}
\subfigure[$\sin^2 2 \theta_{12}$ vs. $|\lambda_1 / \lambda_2|$. ]
{\epsfig{file=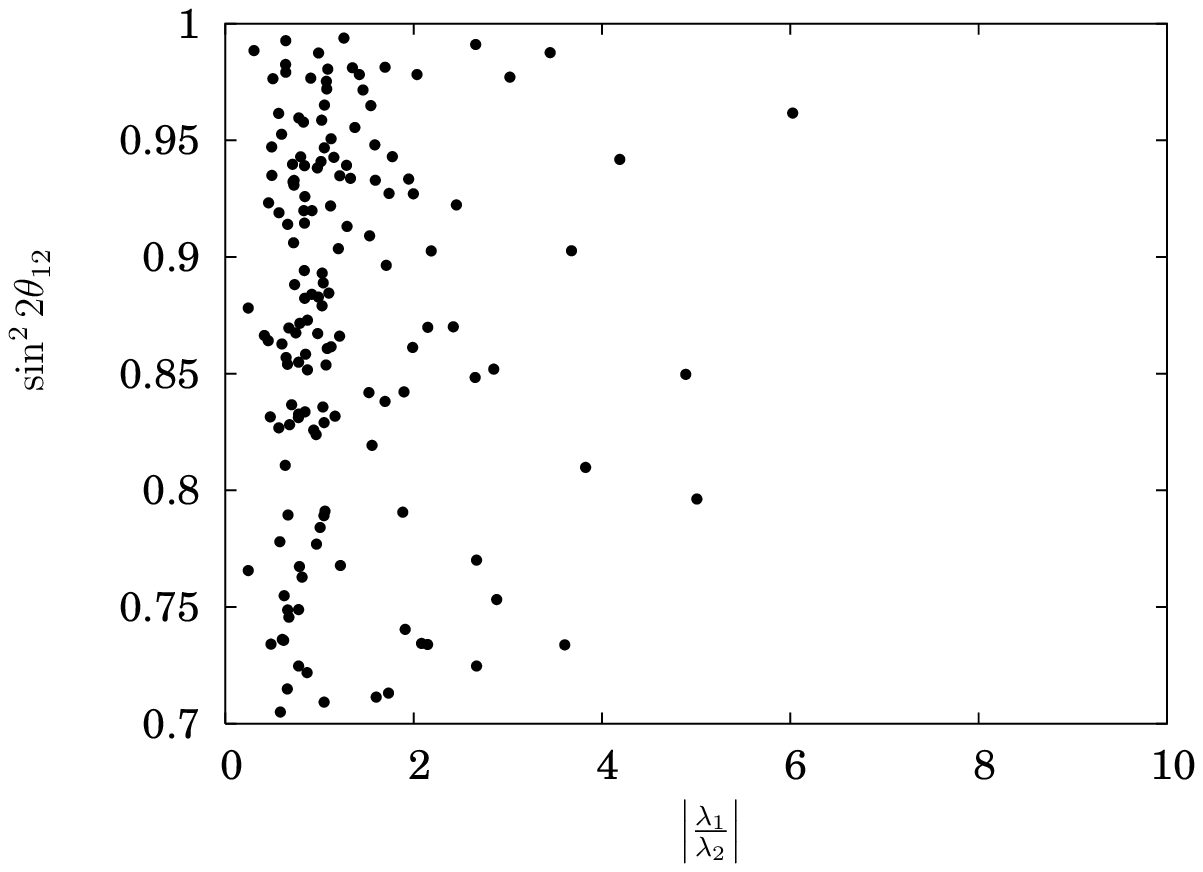,height=3.9cm,width=6.0cm}}
\caption{The correlation between solar neutrino mixing angle and
$|\lambda_1/\lambda_2|$ for solution points with
(a) $\tan \beta = 3 - 15$ and (b) $\tan \beta = 30 - 40$.}
\end{center}
\end{figure}

\section{Collider Signatures}
Such correlations predict some distinctive lepton flavor violating
signatures of the LSP decay in the next linear colliders\cite{desy1},
which can provide a test for the R-parity violation scenario\cite{cjkp}
\cite{valle2}.
If the  stau is the LSP, it is  possible to obtain the information 
on the solar
neutrino mixing angle $\theta_{12}$ since the stau decays into
the final states $l_i \nu$ through the couplings $\lambda_i$.
When $\tan\beta$ is not too large, the relation
$ Br(e\nu) : Br(\mu\nu) : Br(\tau\nu) =
|\lambda_1|^2 : |\lambda_2|^2 : |\lambda_1|^2 + |\lambda_2|^2 $
can be checked to confirm or disprove our scenario.
For large $\tan\beta$, the last relation in the $\tau$ sector is
obscured by the large tau Yukawa coupling effect.  Thus, it will not
be possible to establish the above relation without knowing
other parameters like $\tan\beta$, Higgs masses, etc.
For the case of the neutralino LSP,  its decay to $\nu l_i^\pm \tau^\mp$
can be observed to establish the similar relation as above:
$Br(\nu e^\pm \tau^\mp) : Br(\nu\mu^\pm\tau^\mp) : Br(\nu\tau^\pm\tau^\mp)
= |\lambda_1|^2 : |\lambda_2|^2 : |\lambda_1|^2 + |\lambda_2|^2$.
Again, such a clean relation is invalidated in the $\tau$ sector
for large $\tan\beta$.   If the neutralino is heavier the top quark,
it can decay to $l_i t \bar{b}$ through the couplings $\lambda'_i$,
predicting
$Br(e t \bar{b}) : Br(\mu t \bar{b}) : Br(\tau t \bar{b}) \approx
|\lambda'_1|^2: |\lambda'_2|^2: |\lambda'_3|^2$.
Thus, its measurement can provide information on the
atmospheric neutrino mixing angle.  Let us note that such simple
correlations are valid only for small $\tan\beta \lesssim 15$. We 
present the one solution set as an example.
{\small
\begin{table}
\begin{center}
\begin{tabular}{c|ccc}
\hline
& &  Neutralino LSP &  \\
\hline \hline
&  $\tan\beta=9.7$ &  $ {\rm sgn} (\mu) = -1$ &$\mu=-1045.8$ GeV
\\
&  $ A_0 = 526.5 $GeV  &$ m_0 = 308.5 $ GeV& $M_{1/2} = 947.6$ GeV
\\
\hline
${\lambda}'_i$   &
$9.00\times10^{-6}$ & $6.54\times10^{-5}$ & $-7.61\times10^{-5}$
\\
${\lambda}_i$   &
$6.40\times10^{-5}$ & $7.51\times10^{-5}$ & 0     \\
$\xi_i$ &
$-3.49\times10^{-6}$& $6.97\times10^{-6}$ & $-3.82\times10^{-6}$ \\
$\eta_i$ &
$-3.56\times10^{-6}$& $1.96\times10^{-5}$ & $-2.12\times10^{-5}$
\\
\hline
BR  &  $e$  & $\mu$   & $\tau$   \\
\hline
$\nu t \bar{t}$ &  &97.8\%   &  \\
$ l_i^\pm jj$ & 4.6$\times 10^{-3}$ \% & 1.8$\times 10^{-2}$\% & 5.5$\times
10^{-3}$\% \\
$ l_i^\pm (tb)$ & $ 1.6\times 10^{-3}$\% & $ 3.3 \times 10^{-2}$\%
& $4.3 \times 10^{-2}$\% \\
$ \nu l_i^\pm \tau^\mp$  & $4.3\times 10^{-1}$ \% & $5.9\times 10^{-1}$ \%
& $1.02$ \% \\
\hline
$l_i^{\pm} W^{\mp}$ & $3.39 \times 10^{-3} \%$ & $1.35 \times 10^{-2}$ \% &
$4.05 \times 10^{-3}$ \%  \\
\hline
& $m_{\tilde{\chi}^0_1}$=417.0 GeV
& \hfill $\Gamma=$& $2.04\times 10^{-9}$ GeV  \\
\hline \hline
&  \hspace{1.8cm}$\sigma_{e^+ e^- \rightarrow \tilde{\chi}^0_1
\tilde{\chi}^0_1}$
&$\simeq$ $1.04\times 10^{-1}$ (Pb), & $\sqrt{s}$ = 1 TeV  \\
\hline
\end{tabular}
\end{center}
$$ \begin{array}{l}
(\Delta m^2_{31},~ \Delta m^2_{21})=
(3.6\times10^{-3},~3.2\times10^{-5})~ \mbox{eV}^2  \cr
(\sin^22\theta_{atm},~ \sin^22\theta_{sol},~ \sin^22\theta_{chooz})
=(0.96,~0.98,~0.03) \cr
{\rm The ~decay ~length}~L \simeq 9.66 \times 10^{-6} cm
\end{array} $$
\caption{A solution set with the neutralino LSP which is heavier than
two top quarks.}
\end{table}
}

\section{Summary}
In mSUGRA with R-parity violation, we have found that rather
 large 1-loop corrections are required to satisfy the experimental 
 data of neutrino oscillation. We expect to probe theh mSUGRA withour
 R-parity via LSP decays. We can test the scenarios by scrutinizing the
 various branching ratios in those cases.

\bibliographystyle{plain}

\begin{thebibliography}{99}

\def\plb#1#2#3{Phys.\ Lett.\       {\bf B#1} #2 (#3)}
\def\npb#1#2#3{Nucl.\ Phys.\       {\bf B#1} #2 (#3)}
\def\prd#1#2#3{Phys.\ Rev.\        {\bf D#1} #2 (#3)}
\def\prl#1#2#3{Phys.\ Rev.\ Lett.\ {\bf #1}  #2 (#3)}
\def\mpl#1#2#3{Mod.\ Phys.\ Lett.\ {\bf A#1} #2 (#3)}
\def\rep#1#2#3{Phys.\ Rep.\        {\bf #1}  #2 (#3)}
\def\sci#1#2#3{Science             {\bf #1}  #2 (#3)}
\def\astro#1#2#3{Astrophys.\ J.\   {\bf #1}  #2 (#3)}
\def\epj#1#2#3{Eur.\ Phys.\ J.\   {\bf C#1}  #2 (#3)}
\def\jhep#1#2#3{JHEP              {\bf #1}  #2 (#3)}
\def\ptp#1#2#3{Prog.\ Theor.\ Phys.\ {\bf #1} #2 (#3)}

\bibitem{hall}
L. Hall and Suzuki, \npb{231}{419}{1984}.

\bibitem{chun}
E. J. Chun, {\it et al.}, \npb{544}{89}{1999}.

\bibitem{CK} E.J. Chun and S.K Kang, \prd{61}{075012}{2000}.

\bibitem{desy1} ECFA/DESY LC Physics working group, J. A.
Aguillar-Saavedra et al., hep-ph/0106315.

\bibitem{cjkp}
E.J.~Chun, D.W.~Jung, S.K.~Kang and J.D.~Park,
\prd{66}{073003}{2002};
E.J. Chun, D.W. Jung and S.K. Kang, work in progress.

\bibitem{valle2}
W. Porod, M. Hirsch J.C. Romao and J.W.F. Valle, \prd{63}{115004}{2001}.
A. Bartl, M. Hirsch, T. Kernreiter, W. Porod and J.W.F. Valle,
 hep-ph/0206071.


\end{thebibliography}

\end{document}